\documentclass{aa}  

\usepackage{graphicx}
\usepackage{txfonts}
\usepackage{lipsum}
\usepackage{subcaption}        

\usepackage{lscape}          
                                
\usepackage{placeins}        
                            
\begin{document}

   \title{High-inclination Centaur reservoirs beyond Neptune}

   \author{F. Namouni}

   \institute{Universit\'e C\^ote d'Azur, CNRS, Observatoire de la C\^ote d'Azur, CS 34229, 06304 Nice, France\\
             \email{namouni@oca.eu}}

   \date{Received 8 August 2025 / Accepted 30 October 2025}
 
  \abstract
   { 
 Numerical simulations of the 4.5\,Gyr past evolution of high-inclination Centaurs show that they originated from orbits beyond Neptune that were perpendicular to the Solar System's invariable plane in a region called the polar corridor.  The existence of the polar corridor is  explained by the Tisserand inclination pathways followed by Neptune-crossing objects in the three-body problem.  Recently, a study of Centaur injection in the three-body problem has shown that Neptune-crossing trans-Neptunian objects (TNOs) in the polar corridor with semimajor axes in the range [40:160]\,au have   dynamical times that exceed the Solar System's age, suggesting the possible presence of  long-lived reservoirs that produce high-inclination Centaurs.}
   {We aim to numerically demonstrate the existence of such reservoirs in the Solar System by simulating the TNOs' time-forward evolution under the gravitational perturbations of the giant planets, the Galactic tide, and passing stars. We also aim to assess the efficiency of Centaur injection as a function of the initial  inclination and determine if high-inclination Centaurs can be produced by low-inclination reservoirs.} 
   {The motion of the giant planets, TNOs, and passing stars was simulated using the IAS15 $N$-body numerical integrator of the {\sc REBOUND} package. The Galactic tide was included using the {\sc REBOUND}x package. Two TNO orbit types were considered in the semimajor axis range {[40:140]\,au}: cold TNOs with circular orbits  and hot TNOs with a perihelion range of [32:50]\,au.  The TNO Tisserand parameters, $T$, with respect to Neptune were taken in the range [$-2$:2.8] which corresponds to inclinations far from Neptune in the range [$8^\circ$:$135^\circ$], in order to examine Centaur injection at low and high initial inclinations.}
   {{We find that TNO reservoirs in the semimajor axis range [50:140]\,au are long-lived and their populations peak at $T=0.5$ and $T=-1.5$.} Saturn is found to induce secondary structures in the polar corridor by holding the perihelia of a fraction of high-inclination reservoir material. We find that the Centaur inclination at minimum semimajor axis depends linearly on the Tisserand parameter regardless of the initial semimajor axis.  Its amplitude  shows that low-inclination reservoirs such as the early protoplanetary disk are unlikely to produce high-inclination Centaurs, in contrast to  reservoirs  in the polar corridor. }
   {We have identified the likely location  of the closest reservoirs to Neptune populated by TNOs  captured {in the early Solar System that produce high-inclination Centaurs.  The Legacy Survey of Space and Time of the \textit{Vera Rubin} Observatory will be able to constrain the reservoirs' extent and population size.}}


   \maketitle

\section{Introduction}
Solar System Centaurs with inclinations greater than $60^\circ$ are conventionally known as high-inclination Centaurs. Like low-inclination Centaurs, they originate from planet-crossing orbits beyond Neptune, but unlike low-inclination Centaurs, their numbers cannot be explained by the protoplanetary disk in the early Solar System {\citep{Fernandez16, Fraser22, KaibVolk22}.} High-resolution simulations of the past orbits of 19 real high-inclination Centaurs over 4.5\,Gyr have shown that they tend to be polar with respect to the Solar System's invariable plane, accumulating in the so-called polar corridor that extends beyond Neptune's orbit \citep[hereafter Papers I and II]{NamouniMorais18b,NamouniMorais20b}. The polar corridor's existence is rooted in the conservation of the Tisserand parameter of Neptune-crossing trans-Neptunian objects (TNOs), which ensures that their dynamical evolution follows  Tisserand inclination pathways over gigayear timescales regardless of whether time flows forward or backward \citep[hereafter Paper III]{Namouni22}. 

{The semimajor axis spread of $-4.5$\,Gyr high-inclination Centaur clone orbits in the polar corridor extends from the scattered disk's inner edge to the Oort cloud, indicating a wide region of origin (Papers I and II). The Oort cloud, in particular, is a known possible source \citep{Brasser12,Ito24}. The presence of Oort cloud matter at $-4.5$\,Gyr requires it to have been captured by the Solar System from the material in the stellar cluster of its birth \citep{Fernandez00,Levison10,Brasser12,Jilkova16,Hands19,Kaib19}.  Studies of the relaxation of the early flat protoplanetary disk through the orbital instability phase of the Solar System's planets have demonstrated that the Oort cloud component that was formed through the scattering of disk matter later than $-4.5$\,Gyr cannot explain the numbers of high-inclination Centaurs \citep{Kaib19,Nesvorny19}. Orbits of high-inclination Centaurs may be generated by  hypothetical planets in the outer Solar System from an extended Kuiper belt \citep{Batygin16b,Lykawka23}, but such models neglect the stellar environment of the Solar System known to induce instability over gigayear timescales \citep{Kaib25}.

In this paper we examine a new possible radial origin suggested by the semimajor axis spread of $-4.5$\,Gyr high-inclination Centaur clone orbits in the polar corridor. The $-4.5$\,Gyr semimajor axis probability distribution functions were found to peak outside Neptune's orbit in the semimajor axis range [70:100]\,au (Paper II). These peaks potentially indicate a stable region where TNO reservoirs are currently present in the polar corridor and have been supplying high-inclination Centaurs to the giant planets' domain. As a first step in understanding the dynamical origin of the semimajor axis distribution peaks, we studied the dynamics of Neptune-crossing TNOs  in the  Sun-Neptune-TNO three-body problem \citep[hereafter Paper IV]{Namouni24}. This first step is important as the dynamics of Neptune-crossing TNOs is determined to a great extent by the conservation of the Tisserand parameter that gives rise to the concept of Tisserand inclination pathways.  A Tisserand inclination pathway describes how orbital inclination evolves as Neptune is approached, and how the TNO is possibly injected as a Centaur before it eventually recedes from the planet to the outer Solar System if it does not experience collision or ejection.  The Tisserand inclination pathway depends only on the TNO's Tisserand parameter and the planet's semimajor axis. In particular, it is independent of time.  This makes the Tisserand parameter the perfect tool for probing original inclinations. The study of the TNO injection process in Paper IV, as a function of the Tisserand parameter, revealed the existence of a stable region in the polar corridor with a semimajor axis range of [40:160]\,au where the dynamical time of planet-crossing TNOs exceeds the Solar System's age,  suggesting the possible presence of long-lived Centaur-supplying TNO reservoirs.  The fact that the stable region found in the three-body problem using time-forward integration is consistent with the $-4.5$\,Gyr semimajor axis distribution peaks  obtained in Paper II  using time-backward integration in the presence of the four giant planets is encouraging.  In this study we numerically confirm the existence of this stable region  in a comprehensive time-forward simulation of the Solar System; in doing so, we determined the likely location of the closest reservoirs to Neptune populated by TNOs that were captured in the polar corridor 4.5\,Gyr ago and that produce high-inclination Centaurs.   

In Sect. 2 we describe the initial conditions used in this work and their relationship to the stable region found in Paper IV. In particular, the parameter space is enlarged beyond the three-body stable region's  Tisserand parameter range  and beyond the perihelion range of the past orbits of high-inclination Centaurs studied in Papers I and II.  In Sect. 3 we describe the simulation method used to follow the time-forward evolution of TNO reservoirs subject to the perturbations of the four giant planets, the Galactic tide, and passing stars.  }
The reservoirs' state at the simulation's end is discussed as a function of the initial semimajor axis and Tisserand parameter in Sect. 4. The reservoirs' stability is confirmed, as is Neptune's role in providing Tisserand inclination pathways for the reservoirs' material to follow. It is also found that Saturn produces  secondary structures in the polar corridor as it retains a fraction of TNO perihelia near its orbit.  In Sect. 5 Centaur injection is described. In particular, the number of injected Centaurs is found to depend linearly on the Tisserand parameter.  In Sect. 6 the features of Centaurs orbits are described. {Additional linear relations with respect to the Tisserand parameter are found: for the Centaur minimum semimajor axis, for the Centaur perihelion at minimum semimajor axis, and for the Centaur inclination at minimum semimajor axis.} The third explains why high-inclination Centaurs are unlikely to be produced  from low-inclination reservoirs such as the early protoplanetary disk. Section 7 contains concluding remarks. In the remainder of this paper, the Tisserand parameter is specified with respect to Neptune unless stated otherwise.   

\section{Reservoirs' initial extent}
\begin{figure*}
\centering
\hspace*{-55mm}\includegraphics[width=280mm]{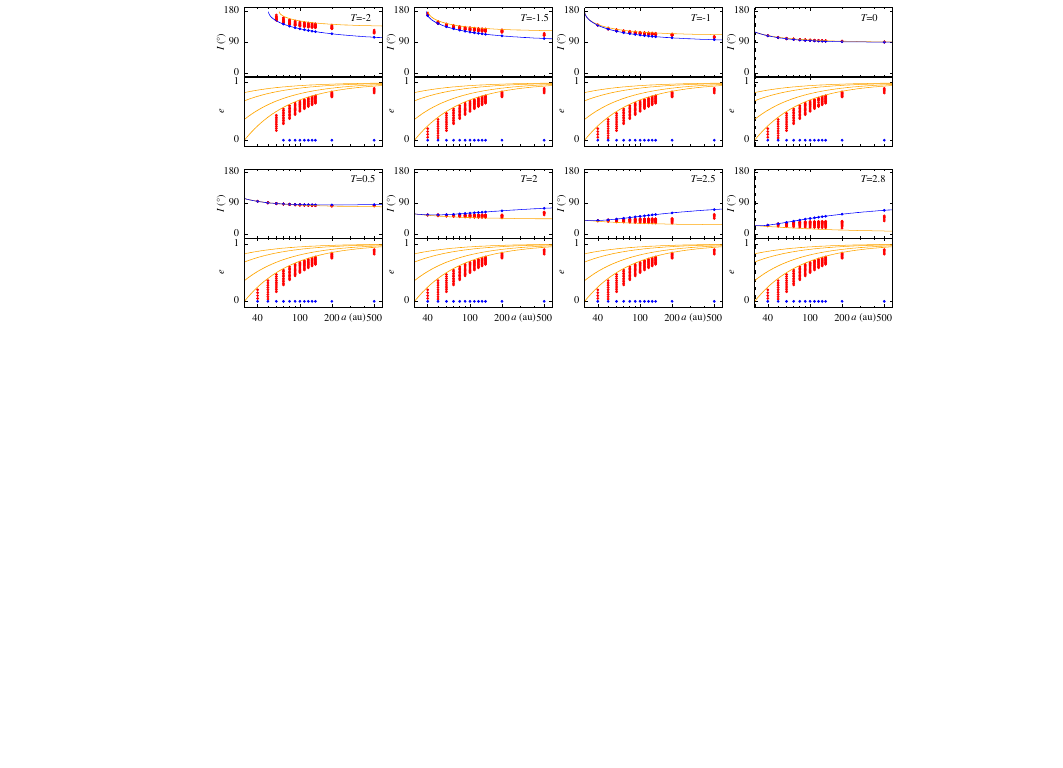}\\[-115mm]
\caption{Initial state of the TNO reservoirs for different Tisserand parameters, $T$, as indicated in the top-right corner of each panel. Hot TNOs  are shown in red and cold TNOs in blue.  The solid orange and blue curves in the inclination panels are, respectively,  the Tisserand inclination pathway (\ref{TissIncP}) and the initial inclination of circular orbits (\ref{Icirc}). In the eccentricity panels, the curves indicate perihelia that match the four giant planets' semimajor axes. }
\label{fig1}
\end{figure*}

{We built the reservoirs' initial state using three results from the 4.5\,Gyr time-backward numerical simulation of 19 real high-inclination Centaurs using $\sim 20$ million clones (Papers I and II). The first result is the Centaurs' past polar state.} Starting with initial inclinations extending over [$62^\circ$:$173^\circ$], the $-4.5$\,Gyr end state inclinations extended over [$60^\circ$:$90^\circ$]  with a standard deviation $\sim 10^\circ$ beyond Neptune's orbit. An analytical analysis of the three-body problem (Paper III) demonstrated that this clustering is the manifestation of the Tisserand parameter's conservation, which makes Centaurs and TNOs follow the Tisserand inclination pathway:
\begin{eqnarray}
 I_{\rm Tiss}(a,T,a_p)&=&\arccos \left(\left[T-\frac{a_p}{a}\right]{\left[4\left(2-\frac{a_p}{a}\right)\right]^{-\frac{1}{2}}}\right),\\ \label{TissIncP}
 \mbox{where}\ \ \ \ T&=&\frac{a_p}{a_0}+2 \left[\frac{a_0(1-e_0^2)}{a_p}\right]^\frac{1}{2} \cos I_0. \label{TissRel}
\end{eqnarray}
Here $e_0$ and $I_0$ are the asteroid's initial eccentricity and inclination and $a_0$ ad $a_p$ are the asteroid's and the planet's semimajor axes. The Tisserand inclination pathway is independent of time and perihelion and defines the polar corridor as the extent in $a$$I$-space between $I_{\rm Tiss}(a,2,a_p)$ and $I_{\rm Tiss}(a,-1,a_p)$.   This corresponds approximately to the inclination range of the $-4.5$\,Gyr orbits of high-inclination Centaurs. 

{The second result is the extent of the $-4.5$\,Gyr semimajor axis distributions peaks} in the region [70:100]\,au. This extent is smaller than the three-body stable region (Paper IV),  [40:160]\,au, where TNO dynamics was followed at Neptune's collision singularity.  The discrepancy in the inner boundary has two reasons. First, the giant planets' perturbations are stronger at the inner edge of the three-body stable region. Secular perturbations from the giant planets, in particular,  increase instability as they can raise significantly the eccentricity of planet-crossing and non-crossing TNOs. Second, the initial conditions at Neptune's singularity include, for a given Tisserand parameter, TNOs with small perihelia that would have crossed the un-modeled orbits of Jupiter, Saturn and Uranus and became unstable. In contrast, Papers I and II indicated that past high-inclination Centaur orbits in the polar corridor did not have arbitrary perihelia but they clustered mainly in the vicinity of Neptune's  orbit.  The discrepancy in the outer boundary is unclear. Despite the proximity of the 100\,au boundary to the location of the outer 1:6 mean motion resonance with Neptune, it was shown in Paper IV that resonances do not fully explain  the stability of high-inclination Neptune-crossing TNOs.

The Tisserand parameter range of the three-body stable region where the dynamical time exceeds 4.5\,Gyr is   [$-1$:0] corresponding to an inclination far from Neptune  $90^\circ\leq I_\infty\leq110^\circ$ where $I_\infty$ is given as (Paper III)\begin{equation}
I_\infty(T)=\arccos (T/\sqrt{8}).\label{Iinfty}
\end{equation}
In Papers III and IV, we demonstrate that a TNO can be injected inside Neptune's orbit only if its Tisserand parameter $T \geq -1$. Once injected, its Centaur inclination becomes retrograde if $T\leq 2$ and remains prograde otherwise. The $T=-1$ limit is specific to the three-body problem and stems from the Tisserand inclination pathway. The  pathway guides the TNO toward the planet up to a reflection semimajor axis that depends on $T$ and Neptune's semimajor axis $a_{\rm N}$ given as
\begin{equation}
a_{\rm refl}=a_{\rm N}[T-2+2\sqrt{3-T}]^{-1}. \label{arefl} 
\end{equation}
The TNO then reverses its radial motion and moves away from the planet. The limit $T=-1$ is when reflection occurs at Neptune's semimajor axis. For $T<-1$, the TNO is reflected outside the planet's orbit and cannot become a Centaur. The addition of Jupiter, Saturn and Neptune to the simulation modifies the three-body injection process and a larger $T$-range is necessary. The Tisserand parameter of a TNO that can cross a planet's orbit must satisfy $-\sqrt{8}\leq T\leq\sqrt{8} \sim 2.82$ (Paper III). To probe the transfer of  prograde orbits to high-inclination retrograde orbits, the reservoirs' Tisserand parameter range is enlarged  to  [$-2$:$2.5$] with a 0.5-step from  the three-body problem  stability range of $[-1$:0]. The additional value  $T=2.8$ ($I_\infty\sim 8^\circ$) was included to provide a means of comparison with low-inclination dynamics.

{We examined two orbit types in the semimajor axis range [40:140]\,au} with a 10\,au-step: hot and cold TNOs. {Hot TNOs are test particles} with eccentric orbits and perihelia in the range [32:50]\,au with a 2 au-step\footnote{For $a=40$\,au, 32\,au$\leq q\leq 38$\,au}. 
These reservoirs could be part of the material captured by the Solar System in its birth cluster. {The perihelion range is based on the third result of the time-backward simulations that showed polar corridor orbits' perihelia clustered at Neptune's orbit. Enlarging the perihelion range beyond Neptune's semimajor axis explores whether the giant planets' combined perturbations produce high-inclination Centaurs from an initial population of non-planet-crossing TNOs.}

 {Cold TNOs are test particles with circular orbits.} These reservoirs have no plausible physical origin. They are used to assess how distant from Neptune can perihelia be before TNOs are no longer injected as Centaurs.   Hot TNO ensembles with ($a=200$\,au and  $32\leq q$(au)$\leq50$ and a 2\,au $q$-step) and ($a=500$\,au,  $50\leq q$(au)$\leq90$ and a 10\,au $q$-step) along with cold ensembles at the same semimajor axes were also simulated to characterize the dynamics far from the reservoirs. 

For each set  ($a$,$T$,$q$), 100 orbits are generated by drawing the orbital angles, mean longitude, longitude of ascending node and argument of perihelion, from a uniform distribution. As the Tisserand inclination pathway is independent of any angles, the 100 orbits evolve on the same pathway in the three-body problem. Furthermore, regardless of the initial $a$ and $q$,  ensembles with the same $T$ value evolve on the same Tisserand inclination pathway.  

Figure\,1 shows the reservoirs' initial extent as a function of eccentricity, inclination and semimajor axis for $T=-2,\ -1.5, \ -1, \ 0, \ 0.5, \ 2, \ 2.5,$ and $2.8$. Hot and cold TNOs are shown in red and blue, respectively. Larger perihelia imply larger (smaller) inclinations than the Tisserand pathway's for (retrograde) prograde motion. This deviation from the inclination pathway increases with the initial semimajor axis and has a limit value, $I_{\rm circ}$,  for circular orbits given as
\begin{equation}
I_{\rm circ}(a,T,a_{\rm N})=\arccos \left(\left[T-\frac{a_{\rm N}}{a}\right]{\left[\frac{4a}{a_{\rm N}}\right]^{-\frac{1}{2}}}\right),\\ \label{Icirc}
\end{equation}
and shown in the inclination panels of Fig. \ref{fig1} as the blue curve along with the Tisserand inclination pathway (\ref{TissIncP}) as the orange curve.

\section{Simulation method}
\begin{figure*}
\centering
\hspace*{-55mm}\includegraphics[width=280mm]{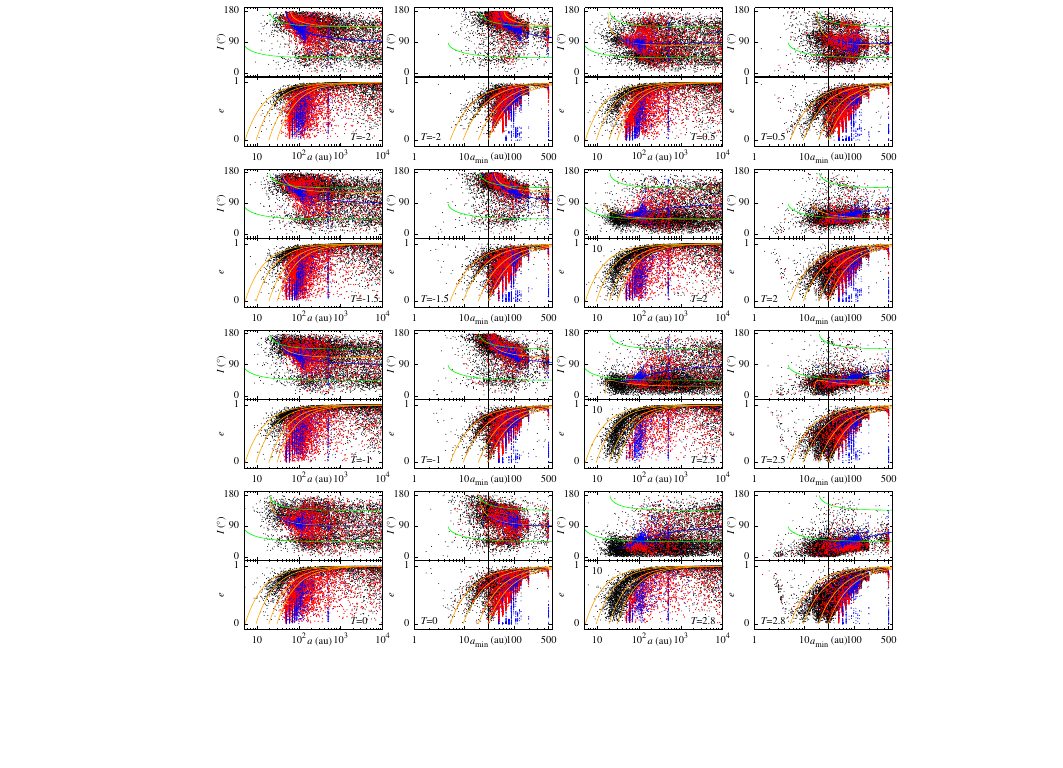}\\[-30mm]
\caption{TNO reservoirs at $4.5\,$Gyr for different Tisserand parameters $(T$). Hot TNOs are shown in red and cold TNOs in blue. Unstable TNOs are shown in black at the last sampling epoch before instability. The first and third columns display the orbital distributions in  the $(a,I)$ and $(a,e)$ planes. The second and fourth columns show $I$, $e,$ and $a$ at minimum semimajor axis, where a vertical black line denotes Neptune's semimajor axis. 
The solid orange and blue curves in the inclination panels are, respectively,  the Tisserand inclination pathway $I_{\rm Tiss}(a,T,a_{\rm N})$ (\ref{TissIncP}) and the circular orbits' initial inclination,  $I_{\rm circ}(a,T,a_{\rm N})$ (\ref{Icirc}), with respect to Neptune. The solid green curves for retrograde and prograde inclinations respectively are the Tisserand inclinations pathways with respect to Saturn $I_{\rm Tiss}(a,-2,a_{\rm S})$ ($135^\circ$ inclination) and $I_{\rm Tiss}(a,2,a_{\rm S})$ ($45^\circ$ inclination). In the eccentricity panels, the curves indicate perihelia that match the four giant planets' semimajor axes. The Tisserand inclination pathway with respect to Neptune is not shown for $T=2.8$ because expression (\ref{TissIncP}) is not valid for $T>2.7$ (see Paper III for details). 
}
\label{fig2}
\end{figure*}

Dynamical evolution is simulated 4.5\,Gyr forward in time using the  IAS15  $N$-body integrator of the {\sc REBOUND} package \citep{Everhart85,Rein15,Rein17}. The simulations include the four giant planets, the TNO ensembles and passing stars. The terrestrial planets' mass was added to the Sun's. The three-dimensional Galactic tide \citep{Heisler86} and relative inclination of the ecliptic and Galactic planes were included using the {\sc REBOUND}x package \citep{Tamayo20}. The Oort constants, $A = 15.3$ km\, s$^{-1}$ kpc$^{-1}$, $B = 11.9$ km\, s$^{-1}$ kpc$^{-1}$ , and star density in the solar neighborhood, $\rho_0 = 0.119 M_\odot$\, pc$^{-3}$, were obtained from \textit{Gaia}'s first data release \citep{Bovy17,Widmark19} and  were used in Papers I and II. 

The list of passing stars was generated using the algorithm of \cite{Heisler87} and the stellar categories of  \cite{Rickman08}  and their corresponding statistics  \citep{Garcia-Sanchez01}. The 66\,230 passing stars are introduced in the simulation according to their Solar System's encounter epoch. Their evolution is integrated within 1\,pc from the Solar System. Multiple stars can be present in the simulation at the same time and therefore interact with one another. The star residency time  within 1\,pc  measured over 4.5\,Gyr is  20\,kyr$\pm$13\,kyr with a maximum value of  415\,kyr. 

{The orbits of 142\,500 TNOs} were integrated and their minimum semimajor axes monitored throughout the simulation. TNOs were removed from the simulation in the case of collision or ejection. 

\section{Reservoirs' state at 4.5\,Gyr}

\begin{figure}
\centering
\hspace*{-15mm}\includegraphics[width=125mm]{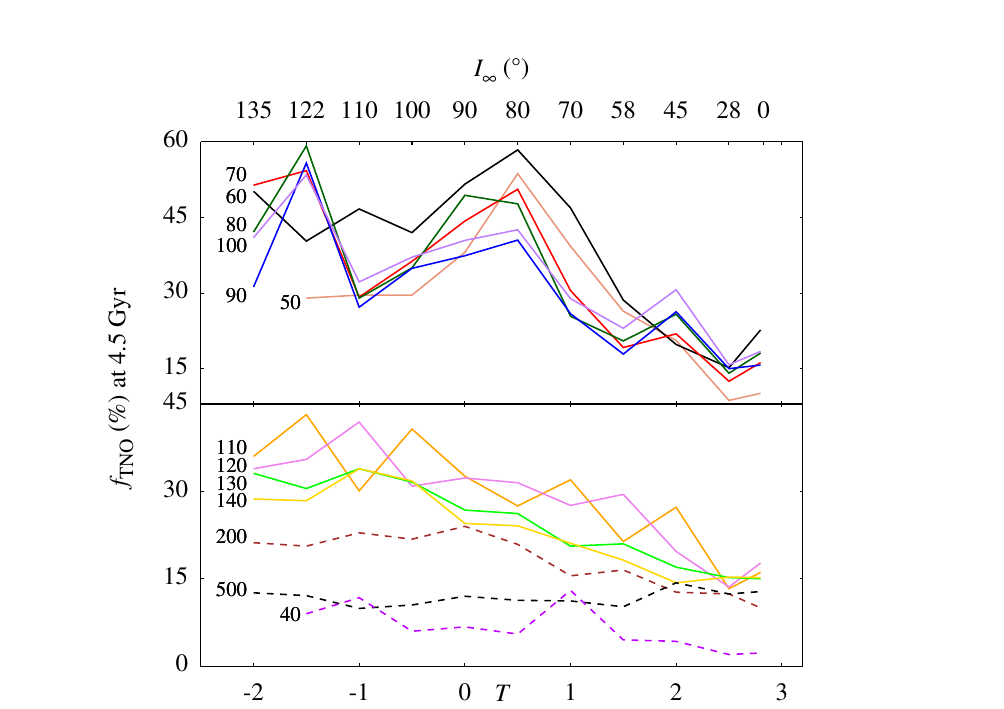}\\[-3mm]
\caption{Fraction of hot TNOs that survived $4.5\,$Gyr as a function of $T$ (and $I_\infty$) for semimajor axes $50\leq a$(au)$\leq 100$ (top panel), $110\leq a$(au)$\leq 500$, and  {$a=40$\,au }(bottom panel). The number next to each curve is the initial semimajor axis.}
\label{fig3}
\end{figure}

\begin{figure}
\centering
\includegraphics[width=\hsize]{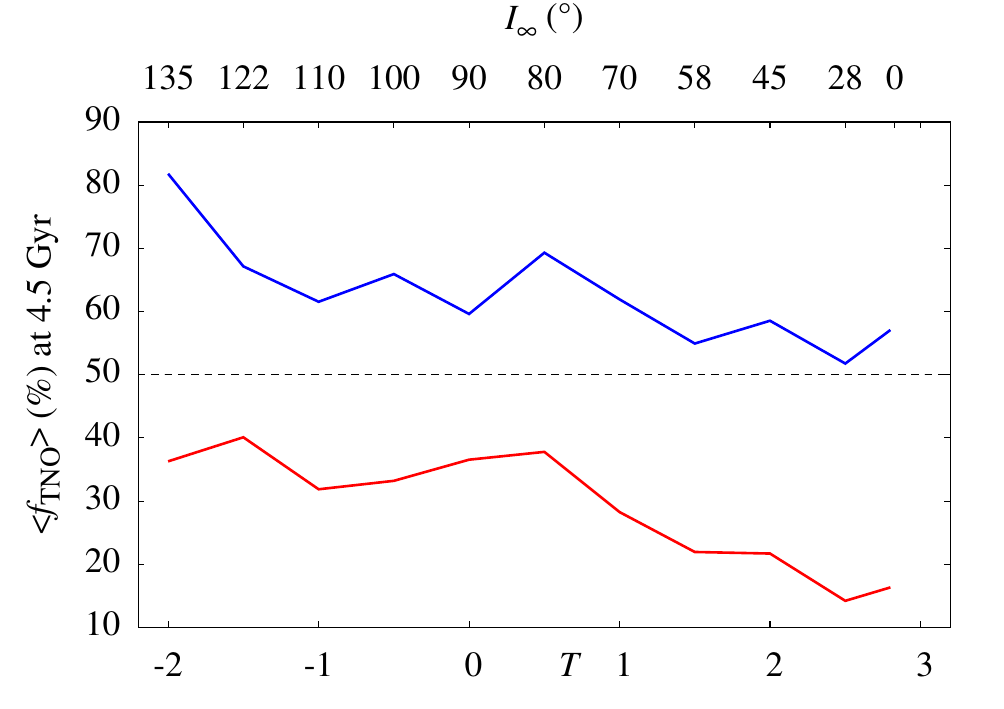}\\[-3mm]
\caption{Average fraction of cold TNOs (in blue) and hot TNOs (in red) that survived $4.5\,$Gyr as a function of $T$ (and $I_\infty$).  }
\label{fig4}
\end{figure}

The end states of the TNO reservoirs in the $Ia$- and $ea$-planes are shown in Fig. \ref{fig2}  for the Tisserand parameters, $T=-2,\ -1.5, \ -1, \ 0, \ 0.5, \ 2, \ 2.5$ and $2.8$ (as in Fig. 1). Hot TNOs are shown in red and cold TNOs in blue. Unstable orbits are shown in black at the last sampled epoch before instability. Also shown are the TNOs' distributions of minimum semimajor axes achieved over 4.5\,Gyr and the corresponding eccentricities and inclinations for all orbital types stable and unstable. 

The reservoirs' semimajor axis extent  is similar to the initial state's. TNOs scattered  by the combined effects of the giant planets, Galactic tide and passing stars are also present far outside the reservoirs' domain. Secular perturbations cause a larger dispersion in eccentricity and inclination for hot TNOs. Cold TNOs remain mainly with low-eccentricity orbits as evidenced by their clustering along the inclination curve of circular orbits (\ref{Icirc}).  As the Tisserand parameter increases toward  the polar corridor boundary's $T=2$, instability is greater and the reservoirs are depleted. Examination of the minimum semimajor axis distributions shows that starting from $T=-2$ minimum perihelia are distributed between Saturn and Neptune but the perihelion range narrows around Neptune's orbit outside the polar corridor as $T$ increases and initial inclinations get smaller. 

The Tisserand inclination pathways with respect to Neptune for the bulk of the reservoirs' material are generally followed for all values of the Tisserand parameter. The pathways are shown in orange in the inclination panels of Fig. \ref{fig2}. A novel process related to Saturn adds secondary inclination pathways that are present for $T\leq2$ and prominent inside the polar corridor. Saturn appears to retain a small fraction of reservoir material with perihelia at Saturn's orbit  whereas their orbits are pushed back outside Neptune's. This material follows two specific  Tisserand inclination pathways with respect to Saturn regardless of the initial Tisserand parameter, $T$. These are  $I_{\rm Tiss}(a,-2,a_{\rm S})$ and   $I_{\rm Tiss}(a,2,a_{\rm S})$ (\ref{TissIncP}) where $a_{\rm S}$ is Saturn's semimajor axis. The pathways correspond  to inclinations far from Neptune $I_\infty=135^\circ$ and $I_\infty=45^\circ$ (\ref{Iinfty}) and appear as the green curves in the inclination panels of Fig. \ref{fig2}. These secondary populations are prominent for $T=0$ and $T=0.5$ in Fig. \ref{fig2} and their density decreases outside the polar corridor as $T$ tends  to $-2$. Investigating Saturn's secondary inclination pathways is beyond the scope of this paper. 

Figure \ref{fig3} shows the ratio of the number of surviving hot TNOs  to their initial number, denoted  $f_{\rm TNO}$, for different initial semimajor axes as a function of the Tisserand parameter. For each semimajor axis, all the different perihelia are included in the statistic -- the perihelion dependence is discussed in the next section. {Four semimajor axis groups are visible.} First, TNOs with initial $a$ in the range [70:100]\,au follow a similar distribution to that of the dynamical time in the three-body problem found in Paper IV (Fig. 7 therein). The dynamical time can be considered a proxy for $f_{\rm TNO}$. The three-body dynamical time versus Tisserand parameter, has three peaks: a prominent one near $T\sim -0.7$, a moderate one near $T\sim 0.8$  and a smaller one near $T\sim 2.2$. The $6$-body plus passing stars configuration of the present framework displaces the three peaks to $T\sim-1.5$  with $f_{\rm TNO} \sim60\%$ for $a=80$\,au, to $T\sim 0.5$ with $f_{\rm TNO} \sim50\%$ for $a=70$\,au and to $T\sim2$  with $f_{\rm TNO}\sim30\%$  for $a=100$\,au.   This group's range of [70:100]\,au is reminiscent of the semimajor axis probability distributions peaks discussed in Sects. 1 and 2. 

{The ensembles with $a=50$\,au and $a=60$\,au form the second group  whose retrograde peak has been all but erased in favor a prominent peak at $T=0.5$ and  $f_{\rm TNO} \sim  60\%$ for $a=60$\,au.
The third group has a semimajor axis  range [110:140]\,au whose $f_{\rm TNO}$  decrease steadily with increasing $T$.  The fourth group is made up of the ensembles of $a=40$\,au, $a=200$\,au and 500\,au that are mainly unstable  confirming the existence of two stability boundaries, an inner one  between $40$\,au and $50$\,au and an outer one between $140$\,au and $200$\,au. } 

The values of  $f_{\rm TNO}$ in the range [50:140]\,au confirm that TNO reservoirs are long-lived in the presence of all giant planets, passing stars and the Galactic tide. In particular, the peaks at $I_\infty=122^\circ$  ($T=-1.5$) and $I_\infty=80^\circ$ ($T=0.5$) single out inclinations that have the largest current populations.  The average  of $f_{\rm TNO}$ over all semimajor axes for hot and cold TNOs, denoted $\langle f_{\rm TNO}\rangle$,  is shown a function of the Tisserand parameter in Fig. \ref{fig4}.  It indicates that cold TNOs on average have a 4.5\,Gyr dynamical time as  $\langle f_{\rm TNO}\rangle>50\%$ unsurprisingly since they mainly retain low-eccentricity orbits. Hot TNOs fractions decrease with $T$ peaking at $T\sim-1.5$ with $\langle f_{\rm TNO}\rangle\sim40\%$,  $T\sim0.5$ with $\langle f_{\rm TNO}\rangle\sim35\%$ and $T\sim2$ with $\langle f_{\rm TNO}\rangle\sim22\%$. 

\section{Centaur injection}
\begin{figure}
\centering
\hspace*{-15mm}\includegraphics[width=125mm]{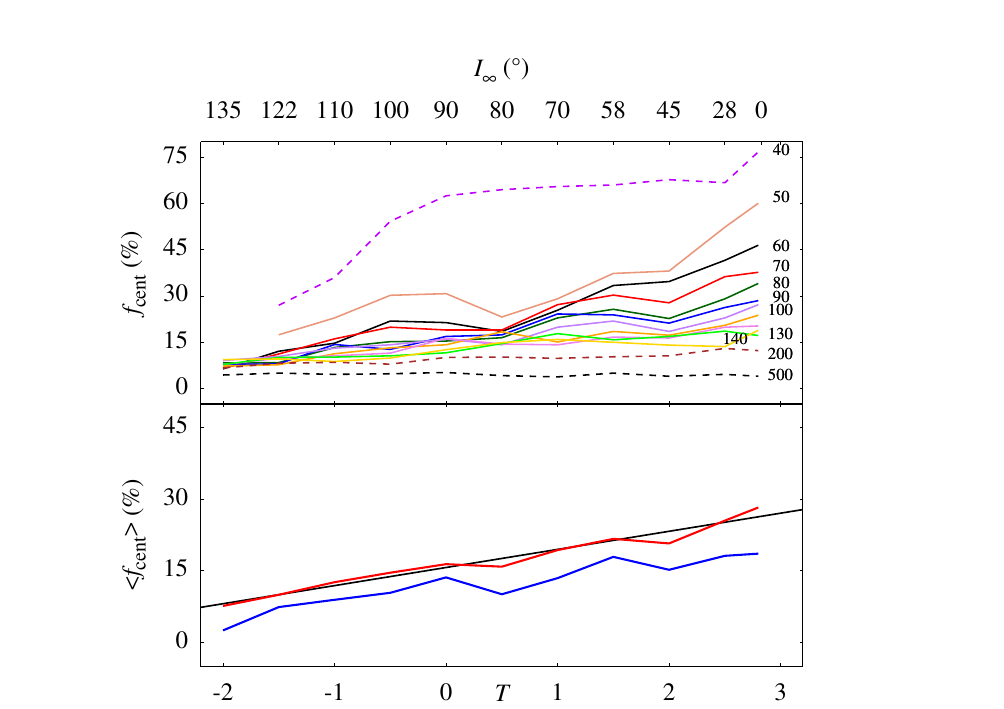}\\[-3mm]
\caption{Fraction of hot TNOs that became Centaurs as a function of $T$ (and $I_\infty$). In the top panel, $f_{\rm cent}$, corresponds to the initial semimajor axis given next to each curve. In the bottom panel, the cold and hot TNO fractions averaged over initial semi-major axes, $\langle f_{\rm cent}\rangle$,   are shown in blue and red, respectively. The black line is $\langle f_{\rm cent}\rangle(\%)=17+4.3T$. }
\label{fig5}
\end{figure}

The ratio of the number of TNOs whose semimajor axes became smaller than Neptune's over 4.5\,Gyr, to the initial TNO number is denoted $f_{\rm cent}(a,T)$. The initial TNO number of each semimajor axis includes the full perihelion range and the Centaur number includes stable and unstable orbits. The top panel of Fig. \ref{fig5} shows  $f_{\rm cent}(a,T)$ for hot TNOs and different semimajor axes as a function of the Tisserand parameter.  The Centaur fraction increases with the Tisserand parameter at a larger rate as the semimajor axis is closer to Neptune. TNOs with low to moderate inclination are injected in greater numbers than high-inclination TNOs as their corresponding reservoirs are depleted because of instability. The hot TNO ensembles with $a=200$\,au and 500\,au contribute smaller Centaur numbers also because of instability. On the other hand, instability for TNOs with $a=40$\,au translates into the largest injection rate because of its proximity to Neptune. For Tisserand parameters $T<-1$, Centaur fraction is finite indicating that the presence of the giant planets and passing stars breaks Neptune's reflecting ability that was evidenced in the three-body problem with the existence of $a_{\rm refl}$ (\ref{arefl}) (Papers III and IV) -- further evidence is given in the next section. 

{We defined $\langle f_{\rm cent}\rangle$ as  the semimajor axis averaged Centaur fraction in the stable region.  The TNO ensembles of $a=40$\,au, 200\,au and 500\,au are excluded from the hot TNO statistics because they lie outside the stability region.  This will be applied in the remainder of the paper when averaging over semimajor axes.
The bottom panel of Fig. \ref{fig5} shows that $\langle f_{\rm cent}\rangle$ of hot and cold TNOs are remarkably similar. The hot TNO } $\langle f_{\rm cent}\rangle$ follows a simple linear expression  that depends on the Tisserand parameter as follows $\langle f_{\rm cent}\rangle(\%)=17+4.3T$.

\begin{figure}
\centering
\hspace*{-2.5mm}\includegraphics[width=103mm]{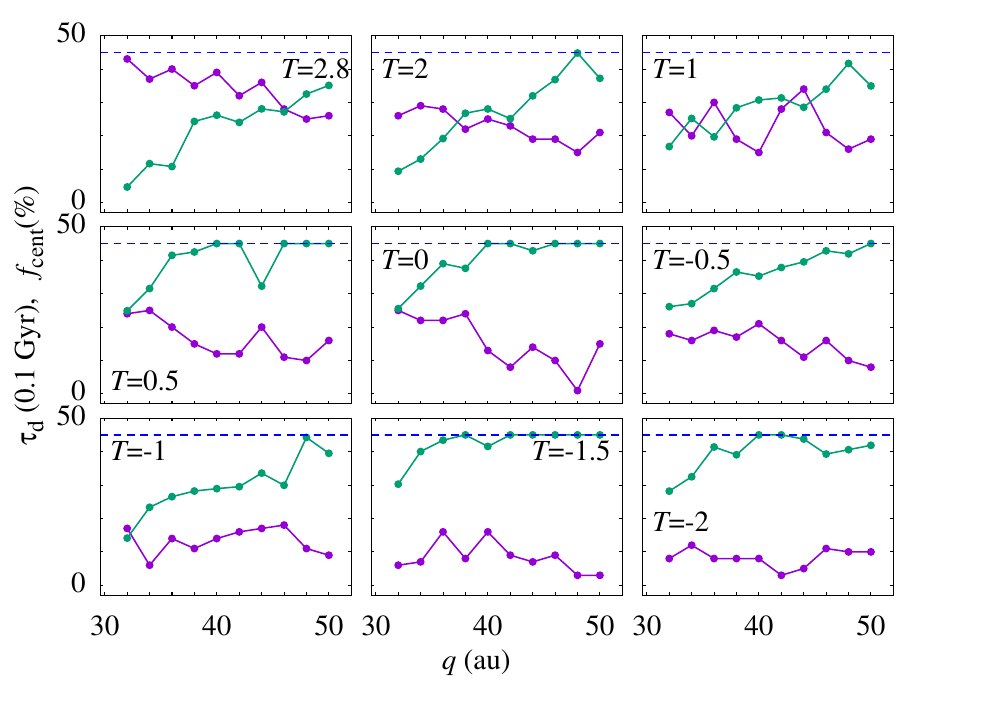}\\[-5mm]
\caption{Dynamical time (green) and Centaur fraction (purple) for initially hot TNOs with an initial $a=80$\,au as a function of perihelion for different Tisserand parameters. The dashed blue line is $\tau_d=4.5$\,Gyr.}
\label{fig6}
\end{figure}

More information on the injection process is found by examining the dependence of the Centaur number on perihelion for hot TNOs and on semimajor axis for cold TNOs. Figure \ref{fig6} shows  non-averaged $f_{\rm cent}(a,T,q)$   along with the corresponding dynamical time, $\tau_d(a,T,q)$, for the ensembles with an initial $a=80$\,au. As stated in Sect. 2, each $(a,T,q)$ ensemble  has 100 TNOs and $f_{\rm cent}(a,T,q)$ is the fraction of those among them that became Centaurs. An  ensemble's dynamical time is its median lifetime. It is meant as a stability indicator and not a dynamical time in the classical sense that is associated with a single orbit and derived from its clones' evolution. As such, the TNOs' dynamical time, $\tau_d$, shown here  is an average of the classical dynamical times of the orbits that make up  the  $(a,T,q)$ ensemble. We find, unsurprisingly, that Centaur injection tends to increase with smaller perihelia for low to moderate inclinations ($T=2.8$). However, the perihelion dependence become shallower for reservoirs inside the polar corridor ($-1\leq T\leq2$) suggesting that it is secular perturbations and not the initial perihelion that control Centaur injection -- see the panel of $T=1$.  The dynamical time is an increasing function of perihelion and reaches 4.5\,Gyr in the polar corridor and for $T>-1$.   Its dependence on the Tisserand parameter inside the polar corridor is unclear but could be depend on the vicinity of $T$ to the boundaries $T=-1$ and $T=2$.  Different initial semimajor axes of hot TNOs in the region [50:140]\,au have similar  $f_{\rm cent}(a,T,q)$ and dynamical times as a function of perihelion. The median and maximum dynamical times for the different perihelia respectively are 3\,Gyr and 4.5\,Gyr, with the latter achieved in the polar corridor.  The median and maximum dynamical times are 2.4\,Gyr and 3.5\,Gyr for  $a=200$\,au and 500\,au, {whereas for $a=40$\,au, they are 0.48\,Gyr and 1.5\,Gyr. }

\begin{figure}
\centering
\hspace*{-2.5mm}\includegraphics[width=103mm]{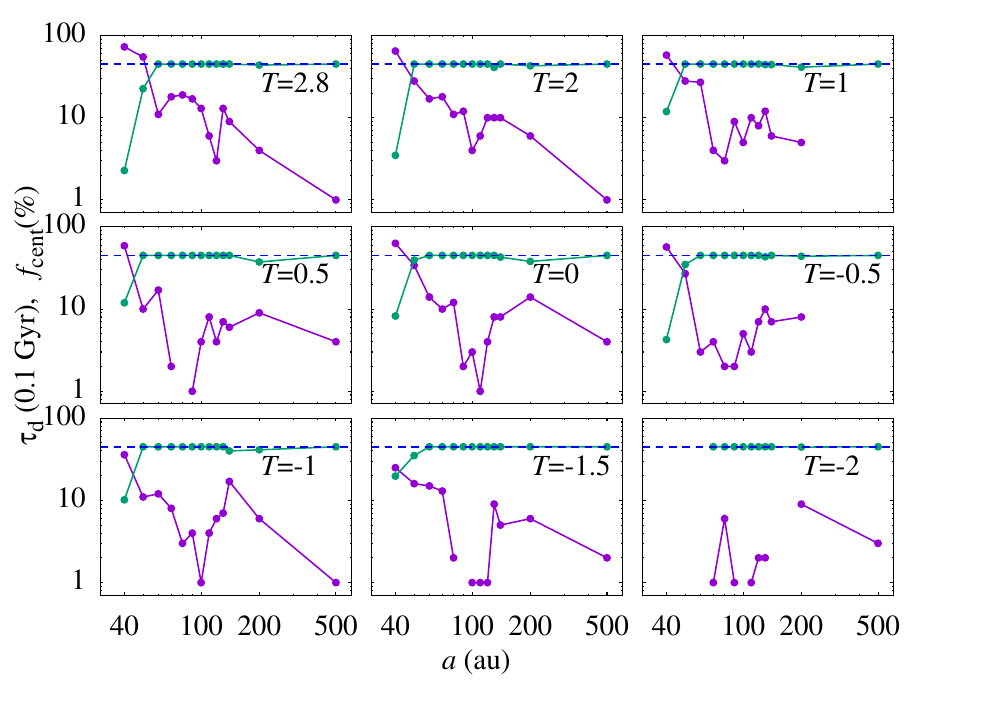}\\[-5mm] 
\caption{Dynamical time (green) and Centaur fraction (purple) for initially cold TNOs as a function of the semimajor axis for different Tisserand parameters. The dashed blue line is $\tau_d=4.5$\,Gyr. The empty semimajor axis ranges indicate the absence of Centaur injection.}
\label{fig7}
\end{figure}

The Centaur fraction, $f_{\rm cent}(a,T,q)$,  and dynamical time of cold TNOs, $\tau_d(a,T,q)$, are  shown in Fig. \ref{fig7}. The number of injected Centaurs decreases exponentially with semimajor axis so that most injected Centaurs come for the smallest semimajor axes. This is visible on the dynamical times  as they reach 4.5\,Gyr as soon as $a\geq 60$\,au regardless of the Tisserand parameter. It is interesting  that regardless of which semimajor axis and perihelion contributes most to Centaur injection, the average Centaur fractions, $\langle f_{\rm cent}\rangle$, of the hot and cold reservoirs are similar (Fig. \ref{fig5} bottom panel). Understanding this similarity's origin may reveal the common dynamical process underlying Centaur injection. 

\section{Injected Centaur orbits}
The analysis of orbit injection in the three-body problem indicates that a Centaur orbit acquires its largest inclination at its minimum semimajor axis as it follows its Tisserand inclination pathway inside the planet's orbit (Paper III). This makes the inclination at minimum semimajor axis a useful diagnostic tool for accessing pre-injection original inclinations far from the planet (\ref{Iinfty}) through the Tisserand parameter.   

\begin{figure}
\centering
\hspace*{-3mm}\includegraphics[width=110mm]{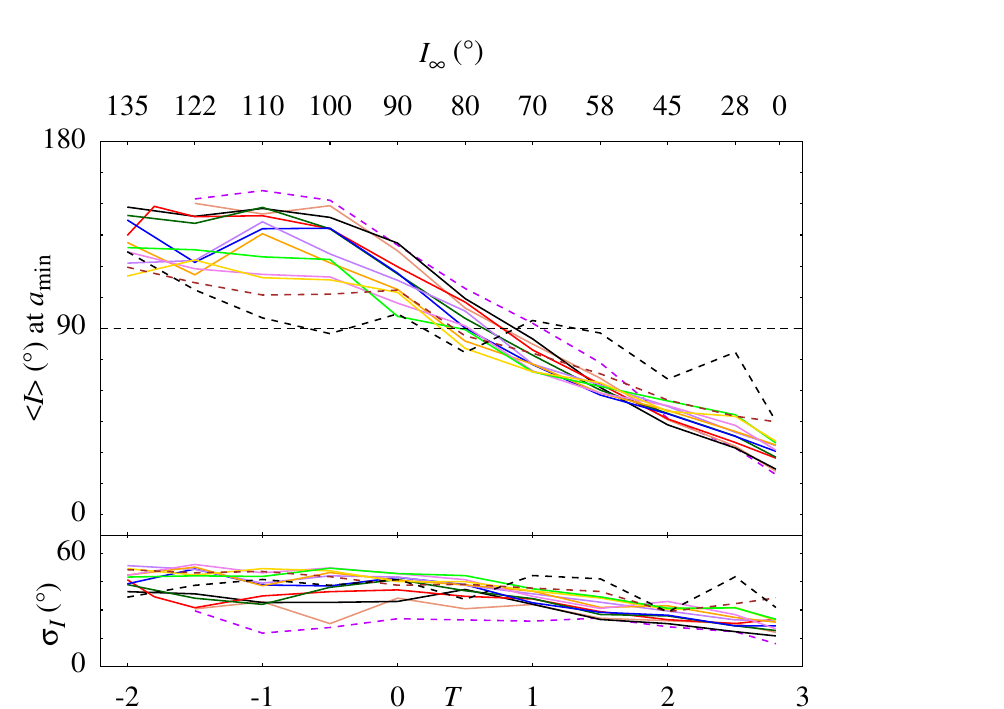}\\[1mm]
\caption{Centaur inclination at minimum semimajor axis  of initially hot TNOs as a function of $T$ (and $I_\infty$) for semimajor axes $50\leq a$ (au) $\leq 140$ (solid curves) inside the stable region, and $a=40,\ 200,\ 500$\,au (dashed curves) outside the stable region. The color codes are those of Figs. \ref{fig3} and \ref{fig5}. The average inclination, $\langle I \rangle$, and standard deviation, $\sigma_I$,  are shown in the top and bottom panels, respectively. }
\label{fig8}
\end{figure}

Figure \ref{fig8} shows the Centaurs' inclination at minimum semimajor axis as a function of the Tisserand parameter for hot TNO reservoirs and the different initial semimajor axes. The statistics for an initial semimajor axis include all initial perihelia. The top panel displays the mean inclinations and the bottom panel the standard deviations. The different initial semimajor axes are not indicated because the curves are close. 

\begin{figure}
\centering
\hspace*{-3mm}\includegraphics[width=95mm]{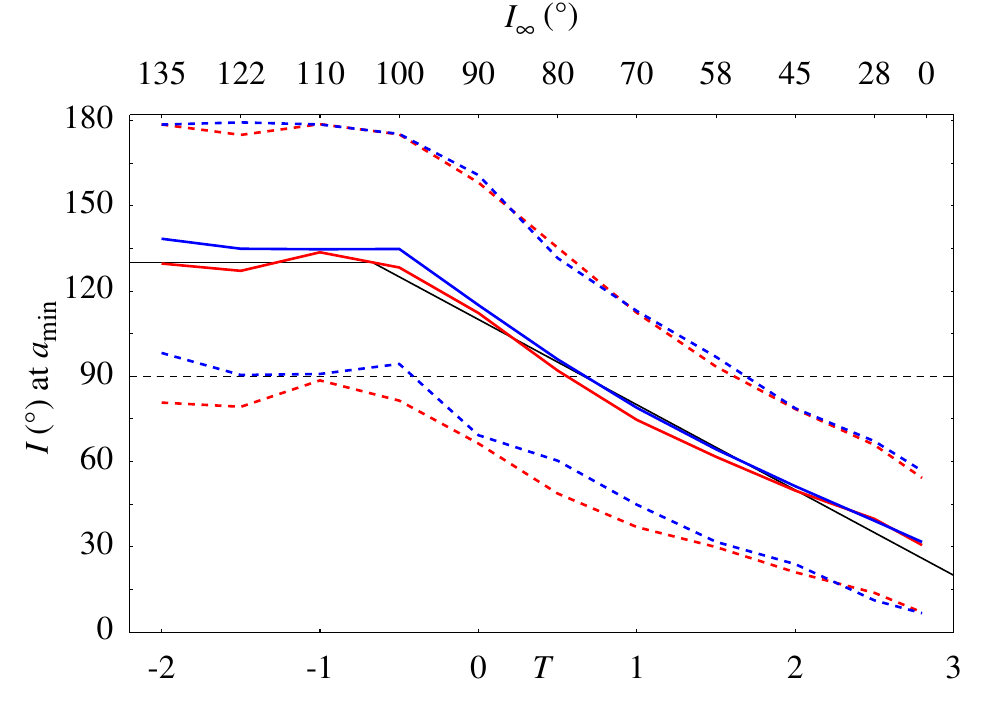}\\[-3mm]
\caption{Semimajor axis-averaged Centaur inclination at minimum semimajor axis  of cold TNOs (blue) and hot TNOs (red) as a function of $T$ (and $I_\infty$).  Dashed blue and red lines are the corresponding standard deviations. The solid black lines are $\langle I\rangle=110^\circ-30^\circ T$ for $T\geq -0.67$ and $\langle I\rangle=135^\circ$ otherwise. }
\label{fig9}
\end{figure}

\begin{figure}
\centering
\hspace*{-3mm}\includegraphics[width=110mm]{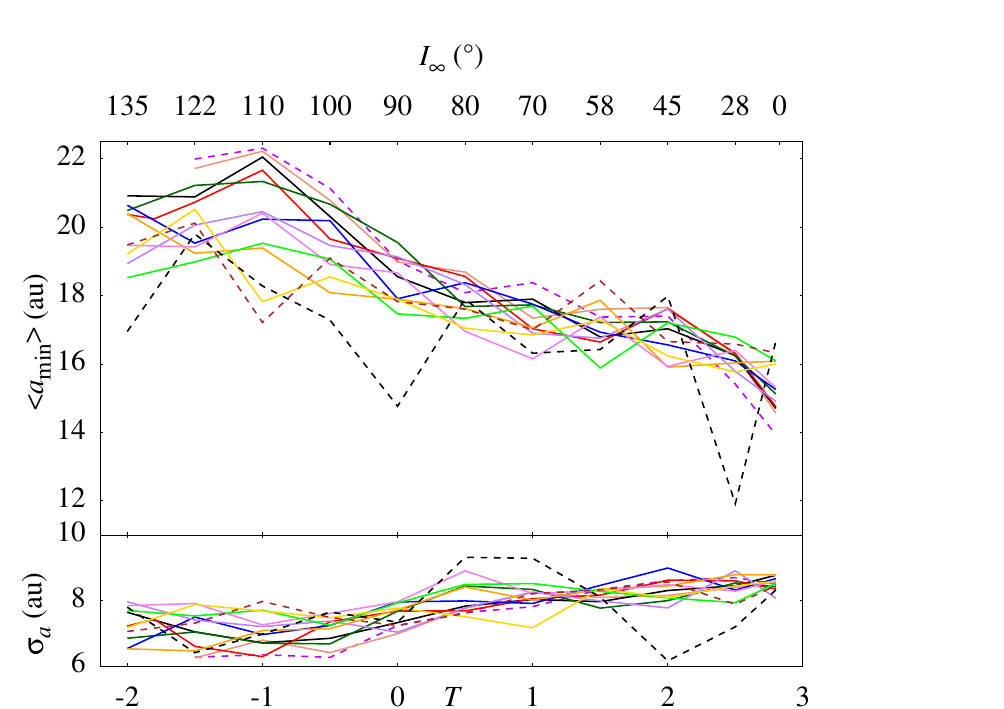}\\[1mm]
\caption{Centaur minimum semimajor axis  of initially hot TNOs as a function of $T$ (and $I_\infty$) for semimajor axes $50\leq a$ (au) $\leq 140$ (solid curves) inside the stable region, and $a=40,\ 200,\ 500$\,au (dashed curves) outside the stable region. The color codes are those of Figs. \ref{fig3} and \ref{fig5}. The average semimajor axis, $\langle a_{\rm min} \rangle$, and standard deviation, $\sigma_a$,  are shown in the top and bottom panels, respectively. }
\label{fig10}
\end{figure}

\begin{figure}
\centering
\hspace*{-3mm}\includegraphics[width=95mm]{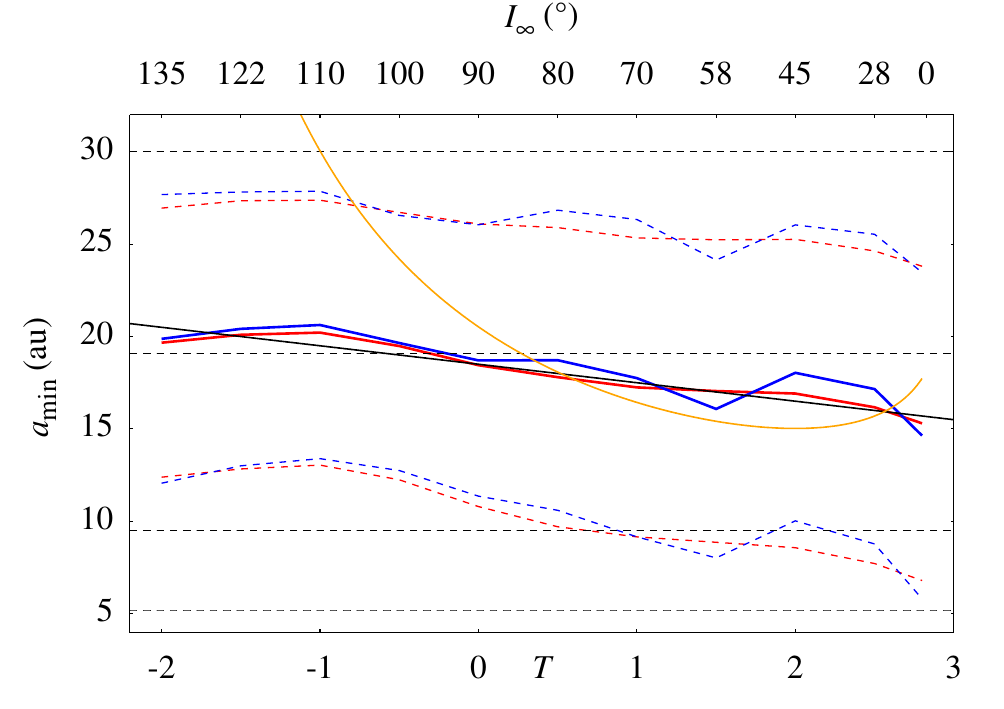}\\[-3mm]
\caption{Average Centaur minimum semimajor axis  of cold TNOs (blue) and hot TNOs (red) as a function of $T$ (and $I_\infty$).  Dashed blue and red curves are the corresponding standard deviations. The solid black line is $\langle a_{\rm min} \rangle$(au) $=18.5- T$. The solid orange curve is the three-body reflection semimajor axis of Neptune (\ref{arefl}). The planets' positions are indicated with dashed black lines.}
\label{fig11}
\end{figure}

The mean inclinations  remarkably share the same functional dependence on the Tisserand parameter regardless of semimajor axis. There is a Tisserand value around $T\sim -0.5$, for each semimajor axis, below which the mean inclination is constant and beyond which inclination decreases linearly with the Tisserand parameter. {The constant inclination value for $T<-0.5$ decreases form $\sim 150^\circ$ for $40$\,au to  $\sim 120^\circ$ for $140$\,au.}

Figure \ref{fig9} shows the semimajor axis-averaged inclinations at minimum semimajor axis {of Centaurs originating in the stable region} as a function of the Tisserand parameter for hot and cold TNO reservoirs along with their standard deviations. The initially hot TNO Centaur-inclination is approximated by $\langle I\rangle=110^\circ-30^\circ T$ for $T\geq -0.67$ and $\langle I\rangle=135^\circ$ otherwise.  The cold TNO Centaur-inclination has a similar dependence. The inclination standard deviation increases steadily from $\sigma_I\sim 25^\circ$ at $T=2.8$ to $\sigma_I\sim 45^\circ$ for $T=-2$.  

The  Centaur inclination at minimum semimajor axis shows that high-inclination Centaurs may not be produced by reservoirs with arbitrary initial inclinations. For instance, a low-inclination reservoir with $T=2.8$ has $\langle I\rangle \sim 30^\circ$. Retrograde Centaurs with inclinations  $100^\circ$  and $155^\circ$ respectively lie at $3 \sigma_I$ and $5\sigma_I$ from the mean value. This explains the inability of relaxation models of the early planar protoplanetary disk to account for the numbers of high-inclination Centaurs. On the other hand, reservoirs in the polar corridor can produce polar and retrograde inclination Centaurs with statistical significance.

The minimum semimajor axes (mean and standard deviation) of hot TNOs that achieved a Centaur orbit are  shown in Fig. \ref{fig10} as a function of the Tisserand parameter for different initial semimajor axes. The minimum semimajor axis is averaged over all initial perihelia. Centaur injection occurs across all values of the Tisserand parameter in contrast to the dynamics of the three-body problem.  The minimum semimajor axes decrease with increasing $T$ in a similar fashion. The large variations of the TNO ensembles with an initial $a=500$\,au comes for the small statistics of injected Centaurs. The  minimum semimajor axes averaged over the different initial semimajor axes  {of Centaurs originating in the stable region} are  shown in Fig. \ref{fig11} for hot and cold TNOs. The two curves are similar and can be approximated by the linear expression $\langle a_{\rm min} \rangle $(au)$=18.5- T$ with a mostly constant standard deviation $\sigma_a\sim8$\,au. The three-body reflection semimajor axis of Neptune (\ref{arefl}) shown in Fig. \ref{fig11} is of the same order as  $\langle a_{\rm min} \rangle $ only for $T>0.4$  as it does not account for the presence of the giant planets.

\begin{figure}
\centering
\hspace*{-3mm}\includegraphics[width=110mm]{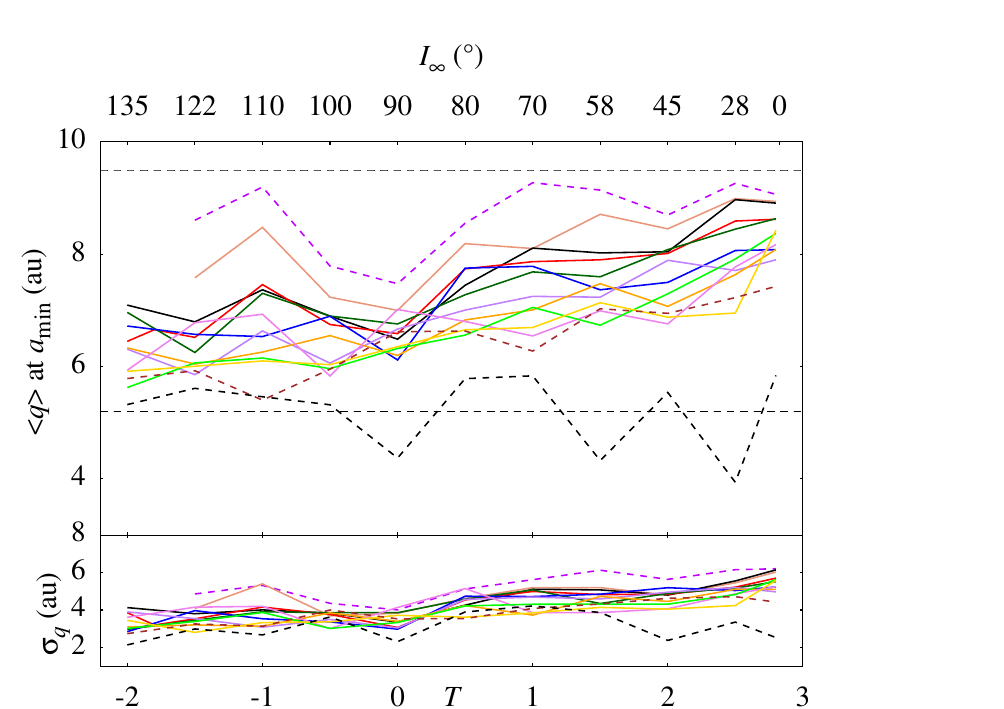}\\[1mm]
\caption{Centaur perihelion at minimum semimajor axis  of initially hot TNOs as a function of $T$ (and $I_\infty$) for semimajor axes $50\leq a$ (au) $\leq 140$ (solid curves) inside the stable region, and $a=40,\ 200,\ 500$\,au (dashed curves) outside the stable region. The color codes are those of Figs. \ref{fig3} and \ref{fig5}.  The average semimajor axis, $\langle q \rangle$, and standard deviation, $\sigma_q$,  are shown in the top and bottom panels, respectively. }
\label{fig12}
\end{figure}

\begin{figure}
\centering
\hspace*{-3mm}\includegraphics[width=95mm]{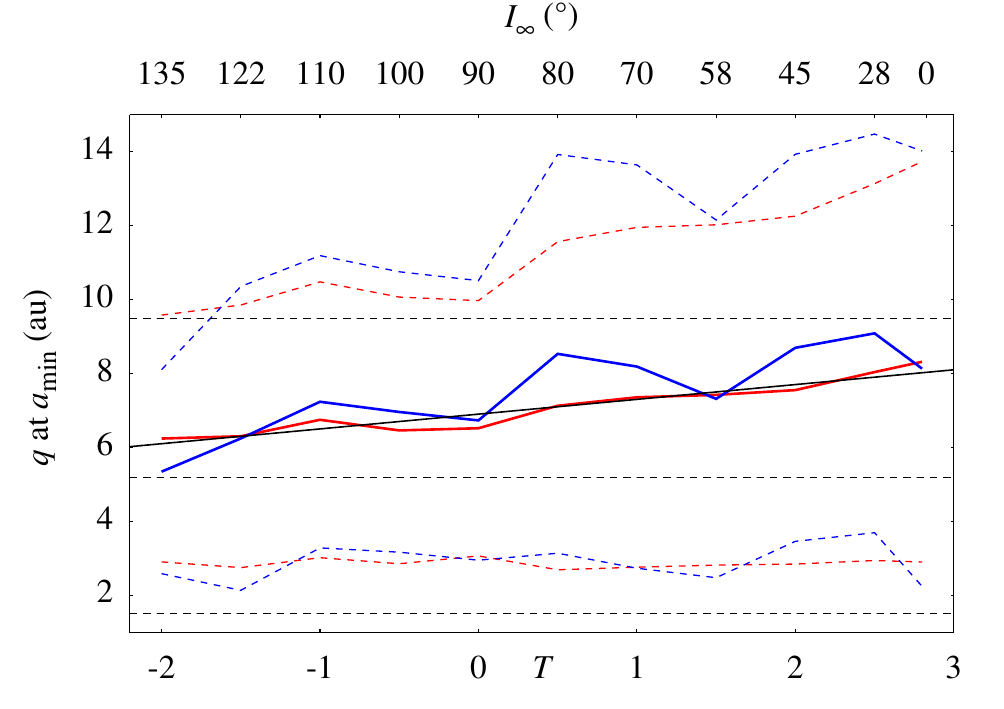}\\[-3mm]
\caption{Average Centaur perihelion at minimum semimajor axis  of cold TNOs (blue) and hot TNOs (red) as a function of $T$ (and $I_\infty$).  Dashed blue and red curves are the corresponding standard deviations. The solid black line is $\langle q\rangle$(au) $=6.9+0.4 T$. The planets' positions are indicated with dashed black lines.}
\label{fig13}
\end{figure}

{The perihelia at minimum semimajor axis  (mean and standard deviation) of hot TNOs that achieved a Centaur orbit are  shown in Fig. \ref{fig12} as a function of the Tisserand parameter for different initial semimajor axes. Mean perihelia and their standard deviations increase with increasing Tisserand parameter (decreasing initial inclination). It is also found that perihelia decrease  with increasing initial semimajor axis.  In particular, hot TNOs with  $a=500$\,au are injected as Centaurs with perihelia around Jupiter's orbit. The semimajor axis averaged Centaur perihelia originating in the stable region are shown in Fig. \ref{fig13}. The hot TNO's mean perihelion as a function of the Tisserand parameter follows the linear relation $\langle q\rangle$(au) $=6.9+0.4 T$. The standard deviation increases from 3\,au at $T=-2$ to 4\,au at $T=2.8$.  
}

\section{Conclusion}
We set out to confirm the possible existence of current TNO reservoirs beyond Neptune's orbit that supply high-inclination Centaurs to the inner Solar System. We did so by simulating  the gravitational perturbations of TNO reservoirs by the four giant planets, passing stars, and the Galactic tide. We based the initial extent of the TNO reservoirs on two previous results. The first is the existence of a stable region in the semimajor range axis [40:160]\,au in the Sun-Neptune-TNO three-body problem (Paper IV). The second is the proximity of past Centaur perihelia to Neptune's orbit found in the 4.5\,Gyr time-backward simulations of 19 real high-inclination Centaurs that involved 20 million clones (Papers I and II). These extents were enlarged to better understand Centaur injection: the perihelion range was enlarged to [32:50]\,au, and a cold population with circular orbits was added to examine if Centaur injection is strongly dependent on perihelion. The Tisserand range was enlarged to [$-2$:2.8], or equivalently to  TNO inclinations far from Neptune in the range $[8^\circ$:$135^\circ],$ to examine Centaur injection beyond the polar corridor's retrograde inclination boundary in the Sun-Neptune-TNO three-body problem, $I_\infty\sim110^\circ$ ($T=-1$), and also at low inclinations of relevance to the relaxation of the early protoplanetary disk,  $I_\infty\sim8^\circ$ ($T=2.8$).

The numerical simulations confirmed the existence of current TNO reservoirs that provide high-inclination Centaurs to the Solar System with  $\sim40$\% of the initial hot TNOs in the polar corridor surviving 4.5\,Gyr (Fig. \ref{fig4}) {in the semimajor axis range [50:140]\,au. It was also found that two specific Tisserand parameter values, $T=0.5$ and $T=-1.5$ (initial inclinations at infinity of $80^\circ$ and $122^\circ$), have the largest current populations (Fig. \ref{fig3})}.  A finer analysis unveiled regions in perihelion space where the dynamical time is 4.5\,Gyr and, hence, surviving hot TNOs constitute more than half the original reservoir (Fig. \ref{fig6}).

The role of the Tisserand inclination pathway with respect to Neptune in the reservoirs' evolution is confirmed (Fig. 2). The presence of all the giant planets modifies the polar corridor's structure as Saturn becomes the main perturber of a fraction of TNO material, adding secondary Tisserand inclination pathways to the polar corridor (Fig. \ref{fig2}). These pathways are defined only by Saturn regardless of the TNOs' initial Tisserand parameters.  Neptune's reflection ability of incoming TNOs is also modified by the giant planets because Centaur injection occurs outside the polar corridor's high-inclination boundary of $T=-1$. Although the presence of more planets causes some significant departures from three-body predictions,  the Tisserand inclination pathway concept  is robust as  TNO dynamics remain based on it. 

The existence of linear relations with respect to the Tisserand parameter that describe the number of injected Centaurs (Fig. \ref{fig5}), the Centaur inclination at minimum semimajor axis (Fig. \ref{fig9}), and the Centaur minimum semimajor axis (Fig. \ref{fig11})  affirms the role of the Tisserand parameter as the main control parameter in TNO dynamics.  These relations were obtained by averaging the orbital elements over initial perihelia and semimajor axes for hot and cold TNOs.  That the two populations follow similar Tisserand parameter-based linear relations indicates that the Centaur injection process is largely independent of the initial perihelia.  {The dependence of the mean Centaur perihelia on the initial semimajor axis shows that the larger the TNO initial semimajor axis, the smaller the perihelion (Fig. \ref{fig12}). Beyond this characterization of the injected Centaur orbits of high-inclination-TNO reservoirs, the linear relations will help us discriminate between different high-inclination Centaur origins, such as the  Oort cloud component formed by the protoplanetary disk and the excitation caused by additional planets beyond Neptune --- provided that Centaur injection is analyzed in terms of the Tisserand parameter using orbital elements at minimum semimajor axis. }

The most compelling result regarding Centaur injection is perhaps how the Centaur inclination at minimum semimajor axis depends on the Tisserand parameter and how it is independent of the initial semimajor axis (Fig. \ref{fig8}). This explains why high-inclination Centaurs are not produced with statistical significance from a low-inclination reservoir such as the early protoplanetary disk, and why matter in the polar corridor can yield Centaurs with retrograde orbits. 

{ Matter in the stable region at $-4.5$\,Gyr is unlikely to originate in the early planetesimal disk because it was flat, and to explain Neptune's current location,  it must not have extended beyond $\sim30$\,au \citep{Gomes04}. The natural source of that matter is the stellar environment of the early Solar System. According to simulations of interstellar matter capture in the Sun's birth cluster,  the inner Oort cloud formed with an inner edge near 100\,au \citep{Brasser12b}, which falls in the middle of the stable region found in this work.  Simulations of star encounters in the presence of protoplanetary disks show that it is possible to capture planetesimals at high inclinations while preserving a star's flat planetesimal disk \citep{Hands19}. After matter exchange in the Solar System's early environment has settled, the stable region of this work  is where the TNO reservoirs can  currently be found nearest to the giant planets. Like  the Trojans asteroids found at Jupiter's stable Lagrange points that are relics of early planet formation, the high-inclination reservoirs are relics of the Solar System's  interactions with the stars of its birth cluster. The Legacy Survey of Space and Time (LSST) of the \textit{Vera Rubin} Observatory will be able to constrain their extent and population size  through its coverage of  high ecliptic latitudes necessary to detect eccentric polar TNOs that spend more time above the ecliptic plane \citep{Kurlander25}.
}

\begin{acknowledgements}
The author thanks the reviewer for their comments. The simulations in this work were performed using HPC resources of the National Facility GENCI-CINES (Grants 2024-AD010415952 and 2025-AD010416591) and on the SIGAMM Cluster hosted at the Observatoire de la C\^ote d'Azur.    
\end{acknowledgements} 

\bibliographystyle{aa}
\bibliography{aa56790-25corr}

\end{document}